\documentclass[journal]{IEEEtran}

\ifCLASSINFOpdf
\else
  \usepackage[dvips]{graphicx}
\fi
\usepackage{url}

\usepackage{color}
\usepackage{graphicx}
\usepackage{amsfonts}
\usepackage{amssymb}
\usepackage{stfloats}
\usepackage{cite}
\usepackage{psfrag}
\usepackage{subfigure}
\usepackage{amsmath}
\usepackage{array}
\usepackage{algpseudocode}
\usepackage{setspace}
\usepackage[english]{babel}
\usepackage{blindtext}
\usepackage{url}
\usepackage{epstopdf}
\usepackage{verbatim}
\usepackage{color}
\usepackage{amsmath}
\usepackage{bm}
\usepackage{multirow}
\usepackage{booktabs}
\usepackage{makecell}
\usepackage{lineno}

\usepackage{caption}

\usepackage{amsthm}
\usepackage{flushend}
\usepackage[ruled,vlined]{algorithm2e}

\begin{document}

\title{Optimizing 5G-Advanced Networks for Time-critical Applications: The Role of L4S }

\author{Guangjin Pan, Shugong Xu, \IEEEmembership{Fellow, IEEE}, Pin Jiang
\thanks{G. Pan and S. Xu are with School of Communication and Information Engineering, Shanghai University, Shanghai, 200444, China (e-mail: \{guangjin\_pan, shugong\}@shu.edu.cn). }
\thanks{P. Jiang is with Wireless X-Labs, Huawei Technologies Company Ltd, Shenzhen, 518000, China (e-mail: jiangpin1@huawei.com).}
\thanks{Corresponding Author: Shugong Xu.}
}

\markboth{IEEE Wireless Communications, Vol. X, No. X, February 2024}
{Shell \MakeLowercase{\textit{et al.}}: Bare Demo of IEEEtran.cls for IEEE Journals}
\maketitle

\begin{abstract}
As 5G networks strive to support advanced time-critical applications, such as immersive Extended Reality (XR), cloud gaming, and autonomous driving, the demand for Real-time Broadband Communication (RTBC) grows. In this article, we present the main mechanisms of Low Latency, Low Loss, and Scalable Throughput (L4S). Subsequently, we investigate the support and challenges of L4S technology in the latest 3GPP 5G-Advanced Release 18 (R18) standard. Our case study, using a prototype system for a real-time communication (RTC) application, demonstrates the superiority of L4S technology. The experimental results show that, compared with the GCC algorithm, the proposed L4S-GCC algorithm can reduce the stalling rate by 1.51\%-2.80\% and increase the bandwidth utilization by 11.4\%-31.4\%. The results emphasize the immense potential of L4S technology in enhancing transmission performance in time-critical applications.
\end{abstract}


\IEEEpeerreviewmaketitle

\section{Introduction}

\IEEEPARstart{W}{ith} the development of technology, immersive and interactive applications have presented a dual demand for transmission speed and latency. Applications such as live streaming, cloud gaming, extended reality (XR), autonomous driving, and metaverse are continuously inspiring people's interest. In order to meet the service requirements of these applications, Real-time Broadband Communication (RTBC) technology is crucial. This includes providing communication services with high bandwidth and low latency to ensure the reliable real-time transmission of real-time media data and real-time sensor data over the network.

However, providing stable streaming transmission with low latency and high bandwidth is not easy. The network requires an appropriate traffic control strategy to prevent congestion. Network transmission latency mainly consists of three components: transmission latency, queue latency, and retransmission latency. The key to providing low-latency transmission lies in minimizing both queuing delay and retransmission delay as much as possible. Traditional transmission mechanisms rely on TCP as the foundation and utilize information such as round trip time (RTT), packet loss, and delay gradients to probe the network’s bandwidth and mitigate congestion dynamically \cite{cite:gcc}. These pieces of information usually come at a cost, and the resulting losses must be remedied. For instance, delay-based congestion control algorithms require noticeable variations in delay to provide congestion information to the sender, resulting in larger queue delays; loss-based congestion control algorithms require retransmission of lost packets due to congestion, leading to even greater retransmission delays. These kinds of end-to-end congestion control schemes manifest as the classic sawtooth waveform shape in TCP, resulting in a long-tailed distribution of delays. Although transport protocols such as QUIC, RIST, and SRT rely on UDP for transmission, they also inherently use costly information for congestion control, and are also unable to provide low-latency transmission. To reduce latency, some protocols such as QUIC have introduced the Explicit Congestion Notification (ECN) field on the Internet Protocol (IP) header for congestion control. However, relying solely on ECN cannot solve the problem of traffic competition, which means that low-latency services for time-sensitive applications cannot be guaranteed.

As mentioned above, these best-effort transmission mechanisms cannot provide satisfactory network services for time-critical applications. In order to solve this problem, a new technology called Low Latency, Low Loss, and Scalable Throughput (L4S) has been defined as a standard by the Internet Engineering Task Force (IETF) \cite{cite:rfc9330}. L4S solves the problem of traffic competition through queue isolation and uses ECN to signal queue delays \cite{cite:rfc9331}. The sender adjusts its transmission rate based on the ECN information, ensuring low queue latency and jitter, thereby enhancing stable flow transmission.

L4S is also recognized as one of the key technologies supporting time-critical applications. As early as 2019, Nokia had already demonstrated L4S networks on a global scale. Apple then added support for L4S in iOS 16. Considering the more challenging demands of low-latency and high-bandwidth transmission in wireless scenarios, in 2022, L4S was adopted in the 3GPP 5G-Advanced Release 18 (R18) to support XR over 5G networks \cite{cite:3GPP23.700-60}. This enables application services and 5G wireless nodes to achieve joint source-channel transmission optimization through the L4S mechanism. Up to today, an increasing number of companies have shown interest in L4S technology \cite{cite:EricssonWhitePaper,cite:NokiaWhitePaper}. Several studies have been conducted to enhance the performance of congestion control algorithms and Active Queue Management (AQM) using L4S and ECN \cite{cite:dualpi2,cite:L4S5G1,cite:L4S_TH}. Nevertheless, there is a notable absence of comprehensive works that offer a detailed introduction to the principles of L4S technology and its incorporation into the 5G standardization process.

In this paper, we first review the principles of L4S technology and summarise its advantages. We also introduced potential applications that can benefit from the L4S technology. Secondly, we introduce the operational mechanism of L4S in 5G-Advanced architecture, and summarize the challenges and opportunities of L4S in 5G-Advanced. Lastly, we further utilized a prototype system to demonstrate the advantages of L4S through a case study based on Real-Time Communication (RTC) applications. Experimental results have shown that L4S technology plays an important role in maintaining low-latency transmission, reducing video stalling rates, and improving bandwidth utilization.

\section{L4S technology}

 In this section, we will first explain the advantages of L4S. Secondly, we introduce its specific technical details to illustrate the reasons for its advantages. Finally, we will introduce the potential applications that L4S can support.

\subsection{Advantages of L4S technology}

L4S aims to address the challenges of providing a high-quality user experience for time-critical applications while also accommodating the growing demand for increased network capacity. It introduces a more responsive and scalable congestion control mechanism that is better suited for real-time applications. The advantages of L4S are introduced as follows.

\subsubsection{Reduced latency}
One key advantage of L4S is its ability to effectively address latency issues. With the assistance of intermediate network nodes, users can seamlessly engage in real-time interaction, and enjoy a smoother and more immersive experience.

\subsubsection{Low packet loss}
L4S reduces packet loss issues caused by congestion by utilizing packet marking instead of dropping. By reducing the probability of packet loss, L4S further minimizes the delay introduced by packet retransmission.

\subsubsection{Throughput optimization}
L4S strives to optimize data throughput without compromising latency. This means that L4S ensures a higher rate of service while maintaining low latency.

In conclusion, L4S provides significant advantages in terms of reduced latency, low packet loss, and optimized throughput, thus enhancing the overall performance and user satisfaction in various real-time applications and services.

\subsection{Architecture and Principles of L4S}

To achieve the aforementioned advantages, L4S realizes low-latency transmission through the following three key mechanisms.

\subsubsection{Queue Isolation} 
Delay-sensitive traffic, such as XR video transmission, often utilizes sensitive flow control strategies that are easily preempted by aggressive flows, resulting in starvation. To mitigate this issue, L4S mandates the implementation of two independent queues in network devices, each with its own AQM policy to separate L4S traffic from classic traffic. This dual-queue mechanism effectively safeguards L4S flows from bandwidth competition, thereby ensuring low latency\cite{cite:L4STMA}.

\subsubsection{ECN-Based Marking Strategy}  L4S redefines the usage of two unused bits in the IP header, known as ECN bits. In L4S, ECN plays two key roles. First, the ECN flag helps network nodes distinguish whether the traffic belongs to classic flows or L4S flows. When a network node detects the L4S ECN codepoint in the IP header is ECT(0) (the value of ECN bits is 10) or Not-ECT (the value of ECN bits is 00), it considers the flow as a classic flow. When it detects an L4S ECN codepoint of ECT(1) (the value of ECN bits is 01), it identifies it as an L4S flow. L4S network nodes use separate queues to schedule traffic for classic flows and L4S flows. Second, when the L4S maintains a relatively short queue length, the ECN codepoint of the packet remains ECT(1). However, when the queue length increases and congestion is imminent, the ECN codepoint of packets is probabilistically marked as CE (the value of ECN bits is 11). L4S AQM immediately responds to queue buildup with an extremely low threshold and simply marks packets that cannot be transmitted in time \cite{cite:rfc9332}. The probability of marking is related to the queue length. Upon receiving the marked packets, the receiver transmits ECN information to the sender in feedback packets, assisting with the sender's rate control.

\subsubsection{Scalable Congestion Control} The congestion control algorithm utilized by L4S is known as Prague Congestion Control, which is a so-called "scalable" congestion control algorithm that uses ECN marking \cite{cite:NokiaWhitePaper}. Based on this congestion control algorithm, the sender can control the application rate with minimal or even no latency or loss, in order to match the network capacity.

\begin{figure}[tb]
\centering 
\includegraphics[height=2.4in,width=3.2in]{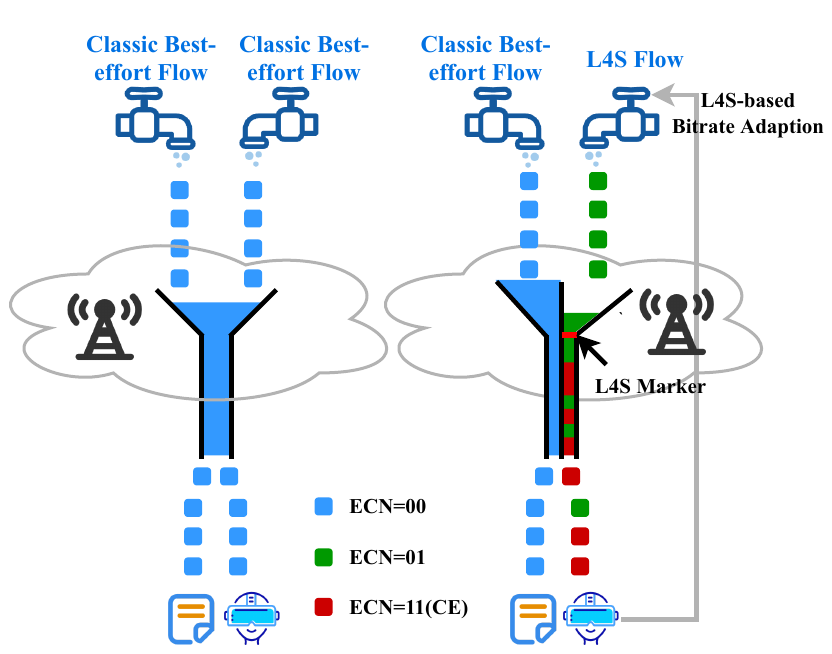} 
\caption{Illustration of L4S Technology Principle. The left figure represents the classic transport model, and the right figure shows the L4S-based transport model.}
\label{fig:L4S} 
\vspace{-3mm}
\end{figure}

The operational mechanism of L4S is displayed in Fig. \ref{fig:L4S}. In the absence of the L4S mechanism (left side of Fig. \ref{fig:L4S}), devices lack cooperation, leading to competitive resource acquisition where each sender adopts aggressive end-to-end congestion control strategies. End-to-end congestion control algorithms typically employ metrics such as packet loss, RTT, and delay gradient to regulate the congestion window. However, the presence of Bufferbloat leads to excessive delay, jitter, or packet loss resulting. Network nodes transmit packets using a best-effort approach, resulting in network congestion and a long-tail distribution of delays. Consequently, this degradation results in latency and overall network performance decline.

With the introduction of the L4S mechanism (right side of Fig. \ref{fig:L4S}), the queue isolation mechanism and ECN-based marking strategy are employed. Firstly, the queue isolation mechanism facilitates the separate transmission of L4S flows and aggressive classic flows, ensuring that L4S flows are not affected by the large queues generated by classic flows. Secondly, the ECN-based marking strategy marks the IP header to CE and provides feedback from the receiver to the sender. This enables the sender to react promptly as the queue shows initial signs of congestion, reducing the bitrate and alleviating queue congestion. Based on intelligent queue management, L4S allows the buffer size to match the network bandwidth, fundamentally solving the impacts caused by large buffers. Furthermore, ECN utilizes marking instead of packet loss as a strategy, preventing L4S flows from experiencing the latency impact brought by retransmissions.

As mentioned above, through mechanisms like Queue Isolation, ECN-Based Marking Strategy, and Scalable Congestion Control, L4S ensures efficient bandwidth utilization and responsive congestion control for time-critical applications such as XR video transmission. The Queue Isolation mechanism protects delay-sensitive traffic from aggressive flows by utilizing independent queues with their own management policies. Additionally, the ECN-Based Marking Strategy allows for traffic differentiation and prompt reaction to congestion signs.

However, the implementation of the L4S also results in increased complexity for the system. Firstly, modifications to existing routing devices are necessary to equip them with queue isolation and ECN marking capabilities. It may complicate the queue management in routers. In addition, the complexity is not confined to hardware alone but also extends to the intricacies of protocol design. The L4S approach needs a joint optimization between the congestion control at the sender and the ECN marking strategies at routing nodes. Consequently, as compared to traditional transmission mechanisms, L4S must consider broader factors to ensure system stability across various aspects, presenting a substantial challenge to L4S system optimization.

\subsection{Potential Applications}

The advantages of L4S enable it to have a significant impact on various applications, such as XR communication, cloud gaming, autonomous vehicles, and so on. Then we will take these as examples to explain its role in detail.

\subsubsection{Immersive XR Communication}
XR applications demand high-bandwidth, low-latency communication for transmitting multimodal data and delivering real-time immersive experiences. To fulfil the user experience demands of XR services, the network should provide a minimum bandwidth of 1 Gbps and low-latency services with no more than 20 ms. L4S technology is well-suited for meeting the communication needs of XR applications. Through network resource scheduling and congestion control, L4S technology can provide higher data transmission throughput while ensuring low latency and low loss, thereby supporting real-time interaction and high-quality experiences in XR applications.

\subsubsection{Cloud Game}
Cloud gaming, based on cloud computing and streaming media technology, relies on real-time game streaming via the internet. To deliver a good user experience and high-performance gaming, low latency and high bandwidth services are essential. L4S technology can offer low-latency communication services, minimizing the delay between users and games, and providing a real-time and smooth gaming experience.

\subsubsection{Autonomous Vehicles}
Low latency and low loss communication are crucial for real-time decision-making and control in autonomous vehicle systems. L4S can improve the performance of vehicle-to-vehicle (V2V) and vehicle-to-infrastructure (V2I) communication, ensuring rapid and reliable data exchange. For low-latency data requirements such as sensor data, traffic information, and navigation commands, L4S reduces network transmission delay, enhancing the responsiveness of autonomous driving systems. Moreover, L4S uses marking instead of packet loss mechanisms to reduce the loss of critical data and enhance the reliability of information transmission in autonomous driving systems. Additionally, L4S enhances network throughput by providing adaptive congestion control and bandwidth allocation mechanisms to support the transmission of multiple data streams.

\subsubsection{Latency-energy-aware Edge Computing}
Edge computing typically needs to provide applications with services that have low latency and are energy-efficient. The use of L4S technology in edge computing not only provides a guarantee for low-latency transmission for edge computing services, supporting the integrity of time-sensitive data, but also enables the determination of transmission priorities according to application demands. By integrating information such as the distribution of node energy consumption, L4S can further optimize edge services that are aware of both delay and energy by expanding the ECN marking scheme, enhancing the overall service quality of the dynamic application placement that is sensitive to latency and energy awareness.

\subsubsection{Digital Twin}
In digital twin systems, it is necessary for the digital twin to stay synchronized with objects in the real world. Therefore, data must be collected quickly and transmitted in real time to the virtual model for updates and analysis. The low-latency transmission provided by L4S technology ensures synchronization between the physical system and the virtual model, which is crucial for precise control and real-time prediction.

\subsubsection{Blockchain}
In order to maintain the consistency of the blockchain, nodes need to continuously synchronize data. L4S can enhance the communication performance within the blockchain network, ensuring fast and reliable data transmission. For example, when conducting cryptocurrency transactions, executing smart contracts, and other operations, L4S technology can improve the responsiveness of the blockchain system by reducing network transmission delay. Additionally, L4S enhances the transmission reliability of information in the blockchain system by using a marking rather than a packet loss mechanism to reduce the loss of important data.

In summary, L4S technology is essential for facilitating high-quality XR communication, cloud gaming, autonomous vehicle systems and other applications, through its ability to provide low-latency, low-loss, and scalable data transmission solutions. Furthermore, L4S also has considerable implications for a wide range of applications including real-time multimedia streaming, remote medical services, telemedicine, and IoT applications, which will not be elaborated here.

\color{black}

\section{L4S in 5G-Advanced}

In wireless systems, due to the constantly changing wireless network environment and the quantity of available resources, applications must adjust their transmission rates accordingly. To improve the transmission efficiency of time-critical services within 5G infrastructures, the 3GPP 5G-Advanced R18 introduces L4S technology to enhance real-time applications such as XR. In this section, we primarily address how the 5G-Advanced architecture supports L4S, and what challenges and opportunities 5G-Advanced faces after introducing L4S. Due to space constraints, more detailed standard details can be found in \cite{cite:3GPP23.700-60,cite:3GPP23.501,cite:3GPP38.300}.

\subsection{How does 5G-Advanced support L4S?}

As a natural multi-queue scheduler, the 5G system can well adapt to L4S technology. It provides strong support for the integration of L4S technology with the 5G system. However, to enable 5G systems to support L4S, two main issues need to be addressed.

\begin{figure}[tb]
\centering 
\includegraphics[scale=0.32]{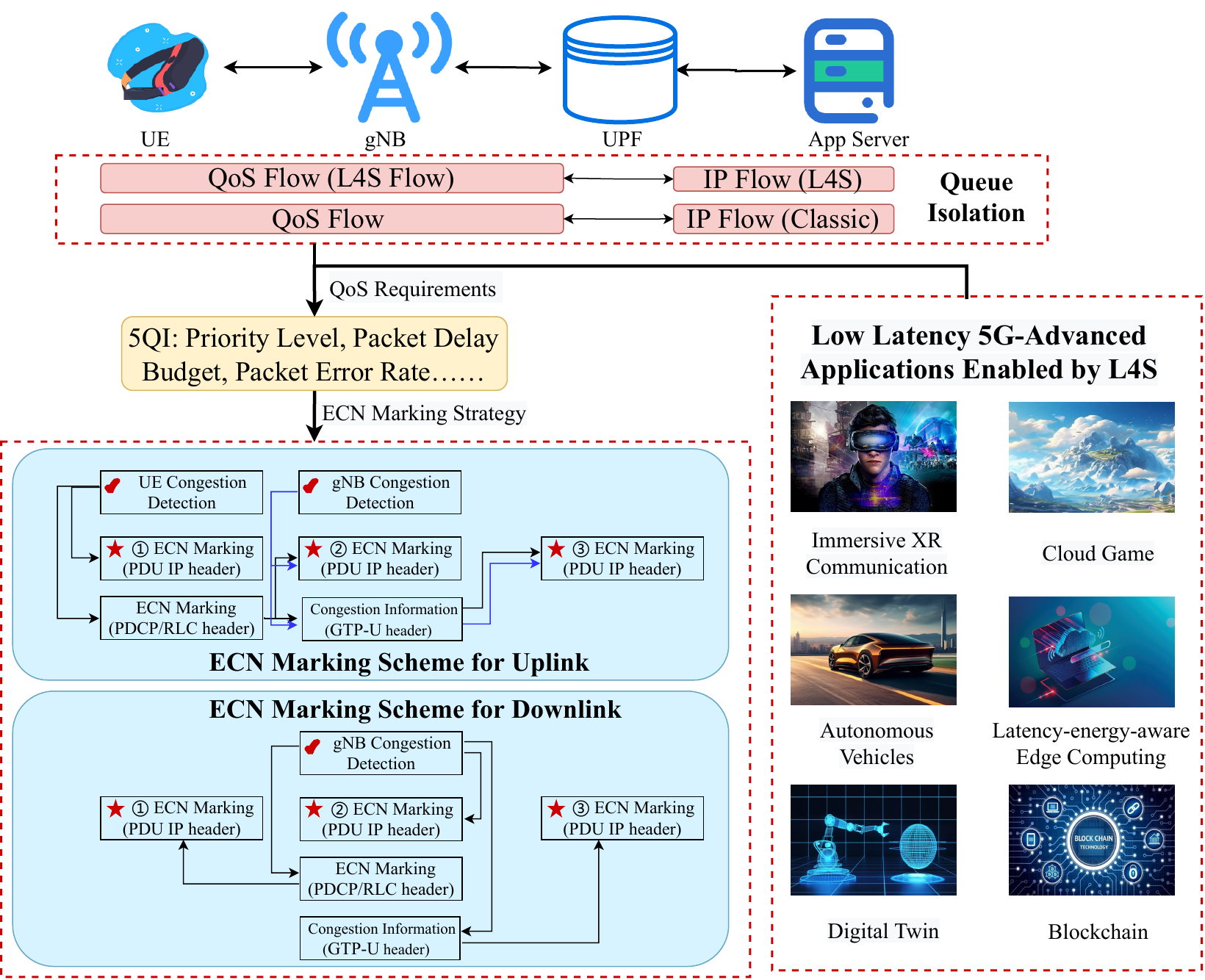} 
\caption{Schematic diagram of L4S in the 5G-Advanced.}
\label{fig:L4S-5GA} 
\vspace{-3mm}
\end{figure}

The first issue is how to combine L4S's queue isolation with 5G's multi-queue. In the 5G system, QoS flows are managed and scheduled to meet varying quality requirements. The differentiation of QoS flows ensures that each flow is treated in accordance with its priority and service requirements. This system inherently provides queue isolation, which is foundational for the support of L4S technology. Therefore, as shown in Fig. \ref{fig:L4S-5GA}, the L4S traffic can be integrated into the 5G QoS framework. Using this intrinsic segregation, A QoS flow designated for L4S traffic can be referred to as L4S flow. The 5G system can distinguish L4S-enabled IP flows by detecting the ECT(1) in the IP header, thereby facilitating queue isolation for L4S traffic. Meanwhile, each QoS flow is assigned a unique 5G QoS Identifier (5QI), which defines the flow's characteristics and requirements, such as bandwidth, latency, and reliability. The definition of each 5QI is tailored to certain application scenarios, allowing for diverse QoS attributes. Accordingly, for L4S traffic, the QoS framework can adapt its ECN marking policy based on the 5QI associated with the flow. This targeted policy enables fine-tuning and optimization of the L4S mechanism for the diversified service scenarios within 5G networks.

The second issue is where to perform ECN marking. The ECN marking refers to L4S routers probabilistically marking the L4S ECN codepoint from ECT(1) to CE. As shown in Fig. \ref{fig:L4S-5GA}, for the uplink, congestion control can be detected either by the user equipment (UE) or by the gNB. When congestion is detected on the UE side, the UE can monitor congestion within the uplink radio bearer to obtain congestion status information. In this case, the UE can directly mark the ECN in the IP Protocol Data Unit (PDU) header, or it can upload the congestion information by marking it in the Packet Data Convergence Protocol (PDCP) or Radio Link Control (RLC) headers, so that the ECN can be marked in the IP PDU header by the gNB or the User Plane Function (UPF). When congestion is detected by the gNB, we can see the blue arrow in Fig. \ref{fig:L4S-5GA}, the gNB can obtain the overall queue established in the UE through buffer status reports and also has a good understanding of the overall congestion situation. The gNB can either directly mark the ECN in the IP PDU header based on the congestion or report congestion information of the QoS flow to the UPF through the General Packet Radio Service in user plane (GTP-U) header to perform ECN bit marking in the UPF.

For the downlink, as shown in Fig. \ref{fig:L4S-5GA}, the gNB performs congestion detection. The gNB can either mark ECN directly in the IP PDU header based on the congestion state, or send congestion information to the UE by marking it on the PDCP or RLC header, with the UE then marking ECN in the IP PDU header. Similarly, the gNB can report congestion information of the QoS flow to the UPF through the gtp-U header to perform ECN bit marking in the UPF, but this will introduce additional latency.

In summary, 5G-Advanced supports L4S through queue isolation by mapping L4S flow to QoS flow 5QI values and implements ECN marking based on congestion detection in both uplink and downlink paths. It enables the 5G-Advanced network to securely expose certain information to applications, aiding real-time applications in transmission.

\subsection{Challenges and Opportunities of L4S in 5G-Advanced.}

Based on the aforementioned discussion, although L4S has already been introduced into the 3GPP standards, system integration also brings some challenges. For example, how to detect congestion in wireless networks? How to design effective ECN marking strategies according to different requirements of different applications? How can bitrate adaptation strategies be coordinated with ECN marking strategies? Therefore, more research is needed to refine L4S implementation in 5G-Advanced systems. A summary of some research topics is as follows.

\subsubsection{Adaptive Congestion Detection Algorithms} Considering the variability in radio conditions and traffic patterns, it is essential to develop effective algorithms that can swiftly detect congestion in both the gNB and UE by employing predictive techniques for channel and traffic conditions. Additionally, it is also important to ensure that these algorithms can work efficiently under different network topologies and conditions.

\subsubsection{ECN Marking Strategy Optimization} When the queue length exceeds a certain threshold, ECN starts marking. Generally speaking, setting a smaller threshold can lead to lower transmission latency \cite{cite:L4S_TH}. However, different applications usually have different delay requirements. Therefore, how to set the ECN marking threshold according to different delay requirements of applications remains a significant challenge. In addition, it is also necessary to explore the optimal ECN marking probability and marking strategies under different congestion conditions in the uplink and downlink. Analyzing the trade-offs between direct marking in the IP PDU header by UE/gNB and the delayed marking through UPF is also crucial for the adoption of L4S technology in 5G.

\subsubsection{L4S-based Congestion Control Algorithms} Design L4S-based congestion control algorithms that collaborate with ECN marking strategies. These congestion control algorithms must effectively utilize congestion signals, such as ECN, to ensure timely rate adaptation without causing unnecessary rate reductions or triggering congestion in the network.

\subsubsection{Wireless Resource Scheduling Optimization} With the assistance of L4S, further optimize the transmission of delay-critical services through methods such as resource allocation. For example, by dynamically adjusting the link transmission rate, resource reservation, network slicing, etc., to ensure the experience of L4S users without excessively impacting other flows.

\subsubsection{Mobility Management} Assessing the implications of UE mobility on L4S performance, and understanding how to maintain consistent ECN marking and congestion detection while supporting handovers between gNBs.

Research based on the aforementioned topics can facilitate the better adaptation of L4S technology to 5G-Advanced networks, thereby further enhancing the quality of experience for real-time applications.

\section{Case Study: L4S for RTC application}
In this section, we introduce L4S technology in RTC video transmission as a case study to illustrate the role of L4S.

\subsection{Prototype deployment}

\begin{figure}[tb]
\centering 
\includegraphics[height=2.8in,width=3.2in]{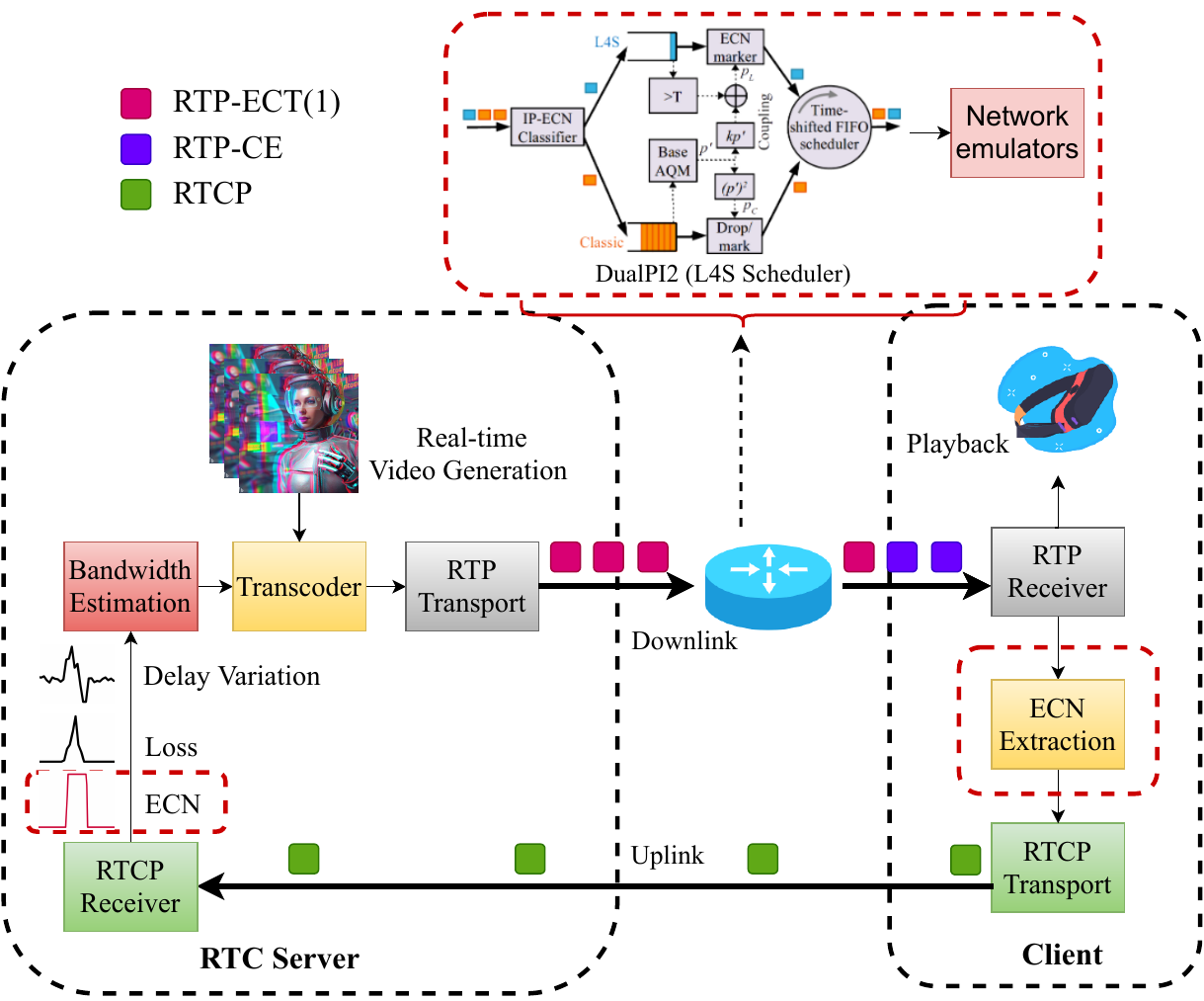} 
\caption{System architecture for L4S-based RTC application.}
\label{fig:L4S-RTC} 
\vspace{-3mm}
\end{figure}

We build an L4S-based RTC transmission prototype system based on WebRTC \cite{cite:webrtc} and DualPI2\cite{cite:dualpi2}. The system architecture is shown in Fig \ref{fig:L4S-RTC}. Specifically, WebRTC is an open standard that supports real-time audio, video, and data transmission between browsers. We select the open-source project Pion, which is developed in the Go language, to implement the WebRTC functionality. Pion complies with the WebRTC standard, provides a wide range of application programming interfaces (APIs), and its modular design facilitates deployment and algorithm verification. Additionally, the DualPI2 is an AQM mechanism implemented as a Linux qdisc, and is one of the open-source solutions for L4S. The DualPI2 scheme can identify L4S and classic flows, and implement ECN marking based on thresholds. Therefore, DualPI2 can realize the functions of the aforementioned L4S routers.

The principle of the original WebRTC bitrate adaptation is as follows. The WebRTC sender continuously transmits RTP packets containing video content in real-time, while the WebRTC receiver collects information about the received RTP packets and provides feedback to the sender through RTCP. Upon receiving the RTCP packet, the WebRTC sender utilizes the packet loss and delay variation to estimate the available bandwidth and adjust the video bitrate accordingly. Congestion control typically defaults to using the GCC algorithm proposed by Google.

The construction topology of the prototype experimental platform is shown in Fig. \ref{fig:L4S-RTC}. In order to introduce L4S into WebRTC, we make three modifications to its transport protocol. Firstly, we enable L4S capability by including the ECT(1) marking in the packets sent by the sender. Secondly, we modified the format of the RTCP packet. During each RTCP cycle, the receiver calculates the number of received ECT(1) packets and CE packets, and provides this information through the RTCP packet as feedback. Third, we optimize the original GCC algorithm in the sender and replace it with L4S-GCC algorithm. Therefore, our experimental prototype consists of the L4S-capable router, WebRTC protocol supporting ECN feedback, and the L4S-GCC congestion control algorithm that supports L4S.

\subsection{Performance Comparison}

The experiment aims to compare the performance of three different approaches,
\begin{itemize}
    \item{GCC}: This approach employs the widely-used GCC algorithm as proposed by Google, employing default configurations. GCC is the default congestion control algorithm in WebRTC \cite{cite:gcc}.
    \item {Sensitive-GCC}: The Sensitive-GCC algorithm utilizes a GCC algorithm with more sensitive parameter settings. The purpose of this algorithm is to detect network congestion more effectively and respond promptly.
    \item {L4S-CC}: The L4S-CC \cite{cite:L4S_TH} algorithm only uses ECN information for bandwidth estimation and bitrate adaptation.
    \item {L4S-GCC}: The L4S-GCC algorithm is an enhanced GCC algorithm based on L4S. L4S-GCC uses L4S to increase the congestion control's responsiveness to bandwidth changes and reinforce the congestion detection mechanism, thus addressing the challenge of under-utilization of bandwidth linked to delay variation.
\end{itemize}

We mainly considered the potential of the L4S mechanism in two aspects compared to non-L4S algorithms. Firstly, whether L4S can respond quickly to bandwidth fluctuations to reduce freezes during video playback. Secondly, we explored whether L4S can effectively cope with scenarios of latency jitter, thereby increasing bandwidth utilization. We assessed the advancements of the L4S technology across three scenarios, 
\begin{itemize}
\item {Case 1: Stable Bottleneck Bandwidth.} The network emulator generates a bandwidth of 3 Mbps without delay jitters.

\item {Case 2: Fluctuating Bottleneck Bandwidth.} The network emulator generates a square waveform pattern of bandwidth, alternating between 2.5 Mbps and 4 Mbps without delay jitters.

\item {Case 3: Bandwidth of Real Trace.} We use a software-defined cellular radio network to acquire real bandwidth traces for experimentation, reflecting the transmission characteristics of 5G networks \cite{cite:tvtbandwidth}. To simulate scenarios where bottleneck bandwidth is less than the video bitrate, ensuring a fair experimental setup, we normalized the bandwidth traces to 0-5 Mbps.

\item {Case 4: Delay Jitter with Probabilistic Delays.} We apply variable delays following chosen probability profiles. Our experimental setup generates packets with delays of 10/12/14/16ms, 10/14/18/22ms, and 10/18/26/34ms, with respective probabilities of 85\%, 10\%, 4\%, and 1\%. The bandwidth is fixed at 5 Mbps, which is sufficient for real-time video transmission at maximum bitrate.

\end{itemize}

\begin{table}[tb]
\centering
\caption{Performance comparison in Case 1, Case 2, and Case 3.}
\begin{tabular}{|>{\centering}p{0.9cm}|>{\centering\arraybackslash}p{2.4cm}|>{\centering\arraybackslash}p{1.7cm}|>{\centering\arraybackslash}p{0.9cm}|>{\centering\arraybackslash}p{0.9cm}|}
\hline
Cases & Algorithms  &RTT (ms), \newline max/min/ave &Stalling Rate& Quality (Mbps)   \\ \hline

\multirow{4}{*}{Case 1} &GCC & 146/15/29& 1.83\%& 2.90\\ \cline{2-5}

&Sensitive-GCC & 122/16/25  & 0.61\%& 2.83\\  \cline{2-5}

&L4S-CC & 58/15/20 & 0.28\%& 2.32\\ 
\cline{2-5}
&L4S-GCC & 79/14/22 & 0.32\%& 2.72\\ \hline

\multirow{4}{*}{Case 2} &GCC & 379/15/42& 3.42\%& 3.26\\ \cline{2-5}

&Sensitive-GCC & 117/14/27  & 0.95\%& 3.19\\ \cline{2-5}

&L4S-CC & 158/13/22 & 0.49\%& 2.43\\ 
\cline{2-5}

&L4S-GCC & 83/14/26 & 0.62\%& 3.01\\ \hline

\multirow{4}{*}{Case 3} &GCC & 256/12/36 & 2.65\%& 2.23\\ \cline{2-5}

&Sensitive-GCC & 127/14/28  & 0.75\%& 2.05\\ \cline{2-5}

&L4S-CC & 161/15/24 & 0.62\%& 1.71\\ 
\cline{2-5}

&L4S-GCC & 263/12/28 & 0.68\%& 1.92\\ \hline

\end{tabular}
\label{table:RTT} 
\end{table}

Our experiment successfully demonstrates the performance advantages of L4S-GCC in the above-mentioned two aspects. First, based on the performance comparison in Table \ref{table:RTT}, it can be observed that our proposed L4S-GCC algorithm outperforms the original GCC algorithm in terms of reducing latency and stalling rate. Although there is a slight decrease in the achieved quality (Mbps), the benefits in terms of reduced delay and stalling are significant. Compared to the GCC, L4S-GCC can reduce the stalling rate by 1.51\%, 2.80\%, and 1.97\%, respectively in three cases.
Compared to the L4S-CC algorithm, the proposed L4S-GCC improves the average bitrate with a lower stalling rate. This is because L4S-CC only relies on ECN for congestion control, which does not allow for rapid bitrate recovery after congestion ends, while L4S-GCC is able to utilize information such as delay gradients to facilitate this process. Results in Table \ref{table:RTT} indicate that the proposed algorithm can effectively reduce buffering and improve the smoothness of the streaming experience.

\begin{table}[tb]
\caption{Performance comparison in Case 4.}
\begin{tabular}{|>{\centering\arraybackslash}p{2.1cm}|>{\centering\arraybackslash}p{2.4cm}|>{\centering\arraybackslash}p{1.3cm}|>{\centering\arraybackslash}p{1.5cm}|}
\hline
Delay Jitter Configuration & Algorithms  &Quality (Mbps) & Bandwidth Utilization  \\ \hline

\multirow{4}{*}{10/12/14/16 ms} & GCC & 4.12 & 82.4\% \\ \cline{2-4}

&Sensitive-GCC & 2.25 & 45.0\%\\  \cline{2-4}

&L4S-CC & 4.03& 80.6\%\\ \cline{2-4}

&L4S-GCC & 4.69 & 93.8\% \\ \hline

\multirow{4}{*}{10/14/18/22 ms} & GCC & 3.73& 74.6\%\\ \cline{2-4}

&Sensitive-GCC & 1.80& 36.0\%\\  \cline{2-4}

&L4S-CC & 4.18& 83.6\%\\ \cline{2-4}

&L4S-GCC & 4.72& 94.4\%\\ \hline

\multirow{4}{*}{10/18/26/34 ms} & GCC & 2.86 & 57.2\%\\ \cline{2-4}

&Sensitive-GCC & 1.93 & 38.6\% \\  \cline{2-4}

&L4S-CC & 3.98 & 79.2\%\\ \cline{2-4}

&L4S-GCC & 4.43& 88.6\%\\ \hline
 
\end{tabular}
\label{table:delayjitter} 
\end{table}

Second, due to the dynamic changes in wireless network channels, as well as resource competition and Hybrid Automatic Repeat Request (HARQ) retransmissions, there is often delay jitter in packet transmission, which can frequently lead to misjudgment of the GCC algorithm based on delay gradient and result in a decrease in bandwidth utilization. The L4S-GCC algorithm directly obtains the queue information through ECN, eliminating the aforementioned interferences and thus beneficially improving bandwidth utilization. Based on the results shown in Table \ref{table:delayjitter}, it is evident that the L4S-GCC algorithm excels in mitigating delay jitter and improving bandwidth utilization. Regardless of the magnitude of delay jitter, this algorithm consistently delivers higher transmission quality and utilizes the available bandwidth more effectively. In all three configurations, the L4S-GCC algorithm achieved higher transmission quality, with average values of 4.69 Mbps, 4.72 Mbps, and 4.43 Mbps, respectively. In comparison, the GCC algorithm achieved lower average values of 4.12 Mbps, 3.73 Mbps, and 2.86 Mbps, while the Sensitive-GCC algorithm had even lower average values of 2.25 Mbps, 1.80 Mbps, and 1.93 Mbps. Compared to the GCC algorithm, L4S-GCC achieves 11.4\%-31.4\% improvement in bandwidth utilization. It is important to note that while the Sensitive-GCC algorithm performs well in Case 1, Case 2 and Case 3, it compromises its ability to withstand delay jitter. These results demonstrate the superior performance of the L4S-GCC algorithm in maintaining higher quality of transmitted data.

\section{Conclusion}

In this paper, we primarily introduced the technical principles of L4S and elaborated on its application scenarios. Regarding the standardization work of L4S in the 5G system, we provided a summary and analyzed the opportunities and challenges of L4S in the 5G-Advanced architecture. Finally, we demonstrated the advantages of L4S technology in guaranteeing low-latency transmission and improving bandwidth utilization through an L4S-based RTC case study.

\bibliographystyle{IEEEtran}
\bibliography{reference}

\end{document}